\documentclass[review]{elsarticle}

\usepackage{lineno,hyperref}
\usepackage[table,xcdraw]{xcolor}
\modulolinenumbers[5]
\usepackage{amsmath}
\journal{Journal of \LaTeX\ Templates}

\begin{document}

\begin{frontmatter}

\title{Mechanical Response of Pentadiamond: A DFT and Molecular Dynamics Study}

\author[1,2]{Levi C. Felix}
\author[1,2]{Raphael M. Tromer}
\address[1]{Applied Physics Department, State University of Campinas, Campinas, SP, Brazil.}
\address[2]{Center for Computational Engineering and Sciences, State University of Campinas, Campinas, SP, 13083-970, Brazil}

\author[3]{Cristiano F. Woellner}
\address[3]{Physics Department, Federal University of Parana, Curitiba-PR, 81531-980, Brazil.}

\author[4]{Chandra S. Tiwary}
\address[4]{Department of Metallurgical and Materials Engineering, Indian Institute of Technology Kharagpur, West Bengal, India}

\author[1,2]{Douglas S. Galvao\corref{mycorrespondingauthor}}
\cortext[mycorrespondingauthor]{Corresponding author: galvao@ifi.unicamp.br}

\begin{abstract}
Pentadiamond is a recently proposed new carbon allotrope consisting of a network of pentagonal rings where both sp$^2$ and sp$^3$ hybridization are present. In this work we investigated the mechanical and electronic properties, as well as, the thermal stability of pentadiamond using DFT and fully atomistic reactive molecular dynamics (MD) simulations. We also investigated its properties beyond the elastic regime for three different deformation modes: compression, tensile and shear. The behavior of pentadiamond under compressive deformation showed strong fluctuations in the atomic positions which are responsible for the strain softening at strains beyond the linear regime, which characterizes the plastic flow. As we increase temperature, as expected, Young's modulus values decrease, but this variation (up to 300 K) is smaller than 10\% (from 347.5 to 313.6 GPa), but the fracture strain is very sensitive, varying from $\sim$44\% at 1K to $\sim$5\% at 300K.
\end{abstract}


\end{frontmatter}


\section{Introduction}

The existence of different possible hybridization of electron orbitals of carbon yields a huge variety of allotropes with notably different physical and chemical properties. Due to this variety, such as sp$^2$ for graphite and sp$^3$ for diamond, many different forms of carbon can be envisioned. For instance, pentagraphene~\cite{zhang_2015,azevedo_2018,desousa_2020,santos_2020} is build by pentagonal tilling patterns where each vertex contains a carbon atom. Schwarzites are another example of materials in which fully sp$^2$ carbon materials exhibit the shape of triply-periodic minimal surfaces mostly formed by hexagons where the curvature is introduced by the presence of heptagonal and/or octagonal rings~\cite{mackay_1991,terrones_1997,phillips_1992,miller_2016,woellner_2018,braun_2018,felix_2019,feng_2020}. The flexibility of carbon bond hybridization even allows the simultaneous presence of sp$^2$ and sp$^3$ in the same material. That is found in amorphous carbon and some other crystalline~\cite{baughman_1993} forms, such as the recently proposed pentadiamond structure. Pentadiamond is a mixed sp$^2$ and sp$^3$ network of pentagonal rings that could be formed by the copolymerization of hydrocarbon molecules spiro[4.4]nona-2,7-diene and [5.5.5.5]fenestratetraene alternating in the vertices of a cubic~\cite{fujii_2020}.

One of the aspects of these carbon structures that has gained much attention is the mechanical properties, such as Young's modulus, Poisson's ratio, tensile strength, and hardness. For instance, the Young's modulus ($Y$) of carbon nanotubes ranges from 0.1 to 1.7~TPa~\cite{poncharal_1999,krishnan_1998} whereas the tensile strength ($TS$) has values in the 100-200 GPa range~\cite{takakura_2019}. Graphene has values of $Y$ and $TS$ of 2~TPa and 130 GPa~\cite{lee_2012,na_website_nd}, respectively. Several studies have searched for materials with hardness which surpasses that of diamond values. The fact that diamond is the hardest material known so far is grounded on well established physical laws~\cite{brazhkin_2019}, which impose many challenges on the synthesis of structures with Young's modulus higher than that of diamond, the so-called ultrahard materials. 

Many works on carbon structures~\cite{blank_1998,serebryanaya_2001,blank_1998a,blank_1998b} have reported hardness values greater than that of the diamond but it was later found to be not real~\cite{brazhkin_2019}. The original work of pentadiamond~\cite{fujii_2020} claimed that it was harder than diamond but it was later retracted due to a mistake in the \textit{ab-initio} calculation of the elastic constants~\cite{fujii_2020a}. Although these exceptional mechanical properties were not confirmed by other independent calculations~\cite{saha_2020,brazhkin_2020} on the same material it is still a novel form of carbon that deserves further investigations. Other properties such as optoelectronics~\cite{tromer_2020} and the possibility of formation of carbon nitrides~\cite{li_2020} were already analyzed. This work aims to investigate the mechanical properties of pentadiamond by calculating the elastic constants and the electron localization function using density functional theory. Also, with molecular dynamics simulations, we investigate the mechanical response beyond the elastic regime for three loading conditions (compression, tensile, and shear) and at different temperatures (under compression) to evaluate their structural stability. To further elucidate the role of bond hybridization on the mechanical behavior of pentadiamond we compared those results to the ones of the fully sp$^3$-hybridized diamond and a fully sp$^2$-like schwarzite. 

\section{Materials and Methods}

\begin{figure}[ht]
    \centering
    \includegraphics[scale=0.4]{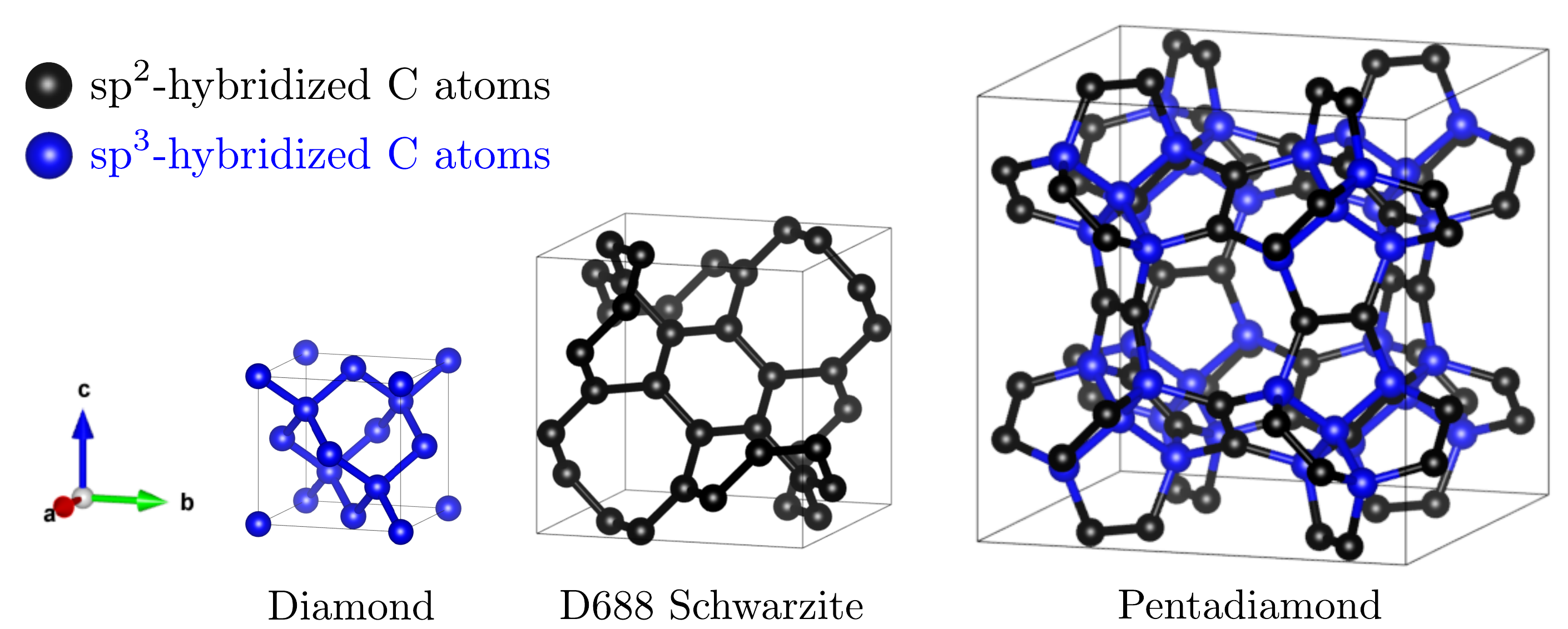}
    \caption{Conventional (cubic) unit cells of diamond (left), D688 schwarzite (middle) and pentadiamond (right), where black atoms are sp$^2$ hybridized and blue atoms are sp$^3$ ones.}
    \label{fig:structures}
\end{figure}

In this work we investigated the electronic and mechanical properties of pentadiamond and compared the results to diamond and a schwarzite structure called D688. Here, D stands for the diamond family of triply-periodic minimal surfaces where each six-membered ring is shared between two eight-membered rings, thus the -688 suffix~\cite{miller_2016}. Diamond and D688 schwarzite are fully sp$^3$ and sp$^2$ hybridized, respectively, and are used to elucidate the role of bond hybridization on the mechanical response of pentadiamond. Figure \ref{fig:structures} presents the conventional (cubic) unit cell of the three materials studied here. The structures of D688 schwarzite and pentadiamond were obtained from the fractional coordinates presented in the works of Terrones \& Terrones~\cite{terrones_2003} and Fujii \textit{et al} \cite{fujii_2020}. Some geometric parameters are shown in Table \ref{tab:structures+dft} such as lattice parameter, fractional coordinates, space group and the number of atoms in the primitive and conventional unit cells. The primitive cells were used in all density functional theory calculations whereas the molecular dynamics simulations were performed from the conventional (cubic) ones.

\begin{table}[ht]
\caption{Structural parameters of diamond, D688 schwarzite and pentadiamond: lattice parameter ($a$), fractional coordinates, space group symmetry and number of atoms in the primitive cell ($n$), where the respective number in the conventional (cubic) cell are given in parentheses. The primitive cells were used in DFT calculations whereas the MD simulations were performed from the cubic ones. The three independent elastic constants $C_{11}$, $C_{12}$ and $C_{44}$. The Voight-Reuss-Hill average of bulk modulus ($B$), Young's modulus ($Y$), shear modulus ($G$) and Poisson's ratio ($\nu$) are also listed.}
\label{tab:structures+dft}
\begin{center}
\resizebox{\columnwidth}{!}{%
\begin{tabular}{c|c|c|c|c|c|c|c|c|c|c|c}
\textbf{Structure} & $\boldsymbol{a}$ \textbf{[\AA]} & \textbf{Fractional coordinates} & \textbf{Space group} & $\boldsymbol{n}$ & $\boldsymbol{C_{11}}$ \textbf{[GPa]} & $\boldsymbol{C_{12}}$ \textbf{[GPa]} & $\boldsymbol{C_{44}}$ \textbf{[GPa]} & $\boldsymbol{B}$ \textbf{[GPa]} & $\boldsymbol{Y}$ \textbf{[GPa]} & $\boldsymbol{G}$ \textbf{[GPa]} & $\boldsymbol{\nu}$ \\ \hline 
Diamond & 3.569 & (0.000,0.000,0.000) & 227 ($Fd\overline{3}m$) & 2 (8) & 1120.05 & 161.88 & 596.31 & 481.27 & 1189.98 & 546.28 & 0.09 \\ \hline
D688 Schwarzite & 6.148 & (0.500, 0.333, 0.666) & 224 ($Pn\overline{3}m$) & 24 (24) & 338.08 & 164.29 & 162.65 & 222.22 & 318.84 & 126.48 & 0.26 \\ \hline
Pentadiamond & 9.195 & \begin{tabular}{@{}c@{}c@{}}(0.250,0.250,0.250) \\ (0.152,0.152,0.152) \\ (0.198, 0.198, 0.000)\end{tabular} & 225 ($Fm\overline{3}m$) & 22 (88) & 531.66 & 99.16 & 143.22 & 243.33 & 411.69 & 169.01 & 0.22 \\ \hline
\end{tabular}
}
\end{center}
\end{table}

\subsection{Density Functional Theory}

We started from their fractional coordinates using their primitive cell configurations and, then, performed geometry optimizations on all structures with variable cell to obtain the lattice parameters (listed on Table \ref{tab:structures+dft}). Then, \textit{ab-initio} calculations of all independent elastic constants and the Electron Localization Function (ELF) were performed with Density Functional Theory (DFT) implemented in Quantum ESPRESSO \cite{giannozzi_2009,giannozzi_2017}. The elastic constants were given by the \texttt{thermo\textunderscore pw} tool~\cite{thermo_pw}. Since all three structures possess cubic symmetry only the three independent elastic constants $C_{11}$, $C_{12}$ and $C_{44}$ were calculated. The macroscopic effective values of bulk ($B$), Young's ($Y$) and shear ($G$) modulus and Poisson's ratio ($\nu$) were obtained from the Voigt-Reuss-Hill average scheme~\cite{toonder_1999} and are listed in the Table \ref{tab:structures+dft}.

It is well known that the elastic moduli are larger in materials with higher binding energy, which corresponds to a significant degree of electron localization on the chemical bonds~\cite{brazhkin_2019}. Thus, the spatial distribution of the electrons are important to qualitatively estimated the mechanical stiffness. The Electron Localization Function (ELF)~\cite{savin_1997} obtained from Quantum ESPRESSO is discussed in the next section in order to analyze the relative mechanical strength among diamond, schwarzite and pentadiamond.

All DFT calculations were carried out within the generalized gradient approximation (GGA), where the exchange-correlation energy term is given by the Perdew–Burke-Ernzerhof (PBE) functional. The interactions between the nucleus and valence electron were described by ultrasoft pseudopotentials. A cutoff energy of $40$ Ry ($\sim 544$ eV) was used in the plane-wave expansion of the wave functions. The lattice parameters and atomic positions of all carbon allotropes studied in this work were fully relaxed until the residual forces on each atom were less then $10^{-4}$ atomic units. The Brillouin zone was sampled using a $8\times 8\times 8$ uniform grid, which was chosen from a convergence test. To check the accuracy of our DFT results, we calculated the band structure and the density of states of the three materials (shown in the Figure S1 of the Supplementary Material) and compared to the literature results~\cite{barnard_2002,huang_1993,fujii_2020}.

\subsection{Molecular Dynamics Simulations}

To evaluate the stability of the mechanical response of pentadiamond beyond the elastic regime, we also carried out fully-atomistic Molecular Dynamics (MD) simulations, as implemented by the open-source code LAMMPS~\cite{plimpton_1995}. The interatomic interaction between carbon atoms was modelled with the AIREBO potential~\cite{stuart_2000}. To better describe the mechanical behavior at large strains we modified the minimum cutoff radius of the force field to 2 \AA~as described by Sherendova \textit{et al}~\cite{shenderova_2000}. Three deformation modes were simulated: compression, tensile and shear. Before all deformations, NPT equilibration MD runs were performed during $100$ ps to eliminate any residual stresses in the structures. Then, the simulation box was correspondingly changed. For compression (tensile) the length of the simulation box along the z direction was decreased (increased) with a strain rate of $10^{-5}$ fs$^{-1}$. To check size convergence, we performed simulations of compression for several cubic supercell sizes and conclude that the one of $5\times 5\times 5$ was sufficiently converged, as show in the Figure S2 of the Supplementary Material. For shear deformations, the $xy$ tilt factor of the simulation box were changed in the same strain rate of the uniaxial (compressive/tensile) deformations. The temperature for all three deformation modes were performed at $1$ K to reduce thermal fluctuations on the stress-strain (SS) curves obtained from MD simulations. A timestep of $0.1$ fs was chosen for all simulations. To analyse the stability of pentadiamond at higher temperatures, we simulated uniaxial compression at higher temperatures of 10, 100 and 300 K. To obtain the SS curves, we calculated the virial stress, given by
\begin{equation}
    \sigma_{ij} = \frac{1}{V}\sum^N_{k=1} m_k v_{ki}v_{kj} + \frac{1}{V}\sum^N_{k=1}r_{ki}f_{kj}, \quad (i,j = x,y,z),
\end{equation}
where the indices $i,j = x,y,z$ correspond to the Cartesian directions, $N$ is the number of atoms in the system, $m_k$ is the atomic mass of the $k$-th atom, $r_{ki}$ ($v_{ki}$) is the atomic position (velocity) of the $k$-th along the $i$ direction, $f_{kj}$ is the corresponding force on the $k$-th atom due to its neighbors and $V$ is the volume of the structure. The compressive strain along the direction of loading is defined as
\begin{equation}
    \varepsilon_c = \frac{L_0 - L}{L_0},
\end{equation}
where $L_0$ is the initial length along $y$ and $L$ is its current value during the compression. The tensile strain is defined as negative value of $\epsilon_c$
\begin{equation}
    \varepsilon_t = \frac{L - L_0}{L_0}.
\end{equation}
For a shear deformation on the $xy$ tilt factor of the simulation box, the shear strain is defined as
\begin{equation}
    \varepsilon_s = \frac{xy}{L_0}.
\end{equation}

The Young's modulus $Y$ defined in the elastic regime is obtained by a linear fitting in the low-strain region (for more details, see the Supplementary Material) where the SS relationship can be approximated by
\begin{equation}
    \sigma_{zz} = Y\varepsilon_c.
\end{equation}

The analysis of the local stress concentration was performed based on the values of the von Mises stress \cite{mises_1913} (which is helpful in the fracture analyses) per atom $k$ calculated by the relation
\begin{equation}\label{eq:atomvonmises}
    \sigma^{k}_{v} = \sqrt{\frac{(\sigma^{k}_{xx} - \sigma^{k}_{yy})^2 + (\sigma^{k}_{yy} - \sigma^{k}_{zz})^2 + (\sigma^{k}_{xx} - \sigma^{k}_{zz})^2 + 6((\sigma^k_{xy})^2+(\sigma^k_{yz})^2+(\sigma^k_{zx})^2)}{2}}.
\end{equation}

Also, Poisson's ratio values were calculated as the negative ratio between the transverse strain $\varepsilon_T$ and the longitudinal one $\varepsilon_L$:
\begin{equation}\label{eq:poisson}
    \nu = -\frac{\varepsilon_T}{\varepsilon_L},
\end{equation}
for both compressive and tensile deformations as a function of the strain.

\section{Results}

\begin{figure}[ht]
    \centering
    \includegraphics[scale=0.3]{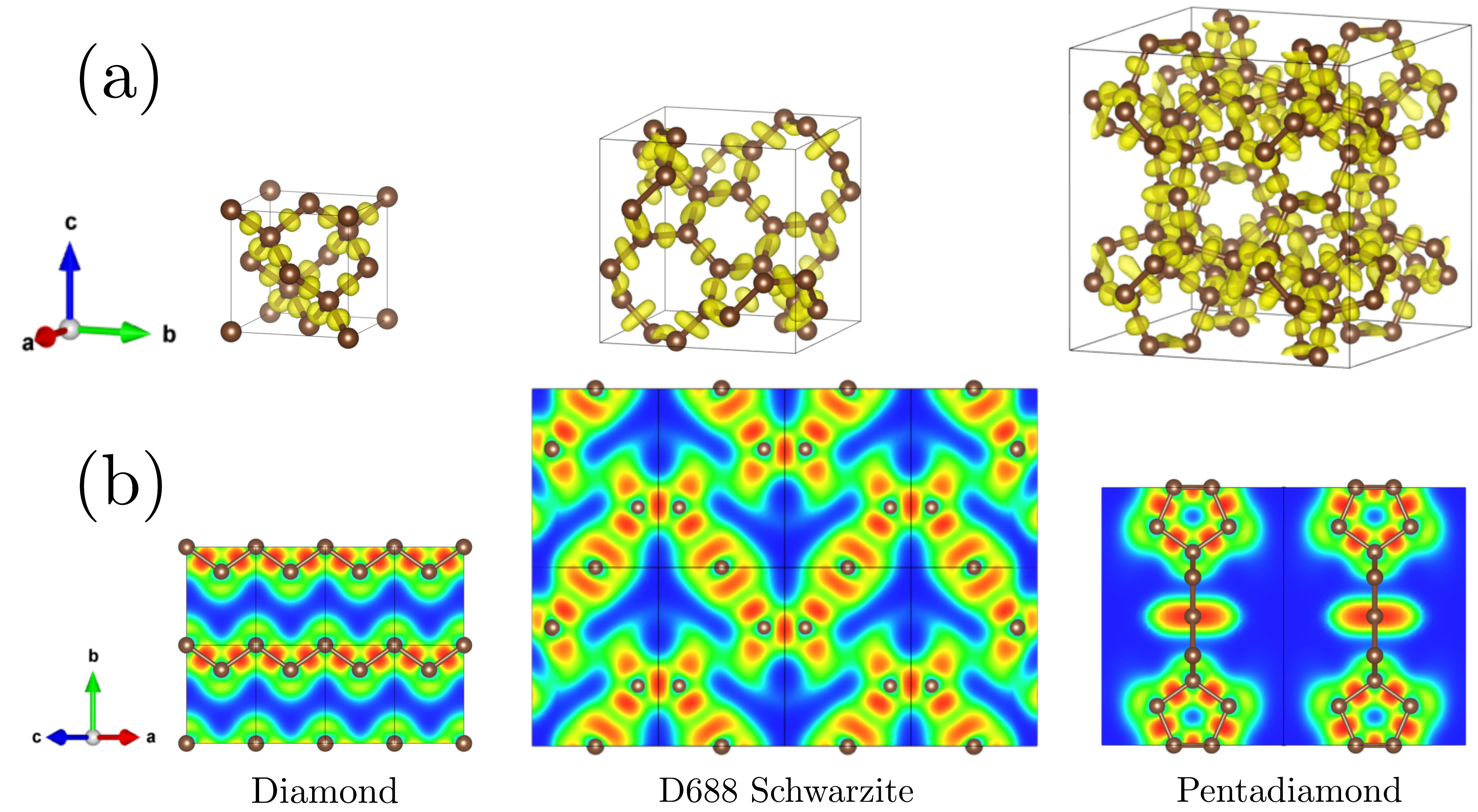}
    \caption{Electron localization function of diamond, D688 schwarzite and pentadiamond as represented by (a) 3D isosurfaces and (b) 2D maps where the normal vector of the planes is parallel to [101] direction. For clarity, we show 2D maps of $2\times 2\times 2$ supercells for diamond and schwarzite. The same isovalue was used on the generation of all the isosurfaces shown in (a).}
    \label{fig:elfmaps}
\end{figure}

The values of the elastic constants shown in Table \ref{tab:structures+dft} show good agreement with previous calculations~\cite{brazhkin_2020,saha_2020}. This validates our methodology to analyse the ELF maps of various carbon materials, shown in Figure \ref{fig:elfmaps}. It can be seen that the electrons are more localized in space in diamond than in schwarzite since the yellow surfaces Figure \ref{fig:elfmaps}(a) and the red spots on the ELF maps shown in Figure \ref{fig:elfmaps}(b) are more concentrated close to the covalent bonds, whereas in schwarzites they are more spread out. The electron distribution in pentadiamond is a mixture of localized and spread out charges, since it possess both sp$^2$ and sp$^3$ hybridized bonds, which can be better observed in Figure \ref{fig:elfmaps}(b). This shows that diamond has a higher bulk modulus than schwarzite and pentadiamond because the electrons are more localized and, thus, has a higher elastic moduli value. This also indicates that pentadiamond possesses an intermediate value between those of diamond and schwarzite.

\begin{figure}[ht]
    \centering
    \includegraphics[scale=0.1]{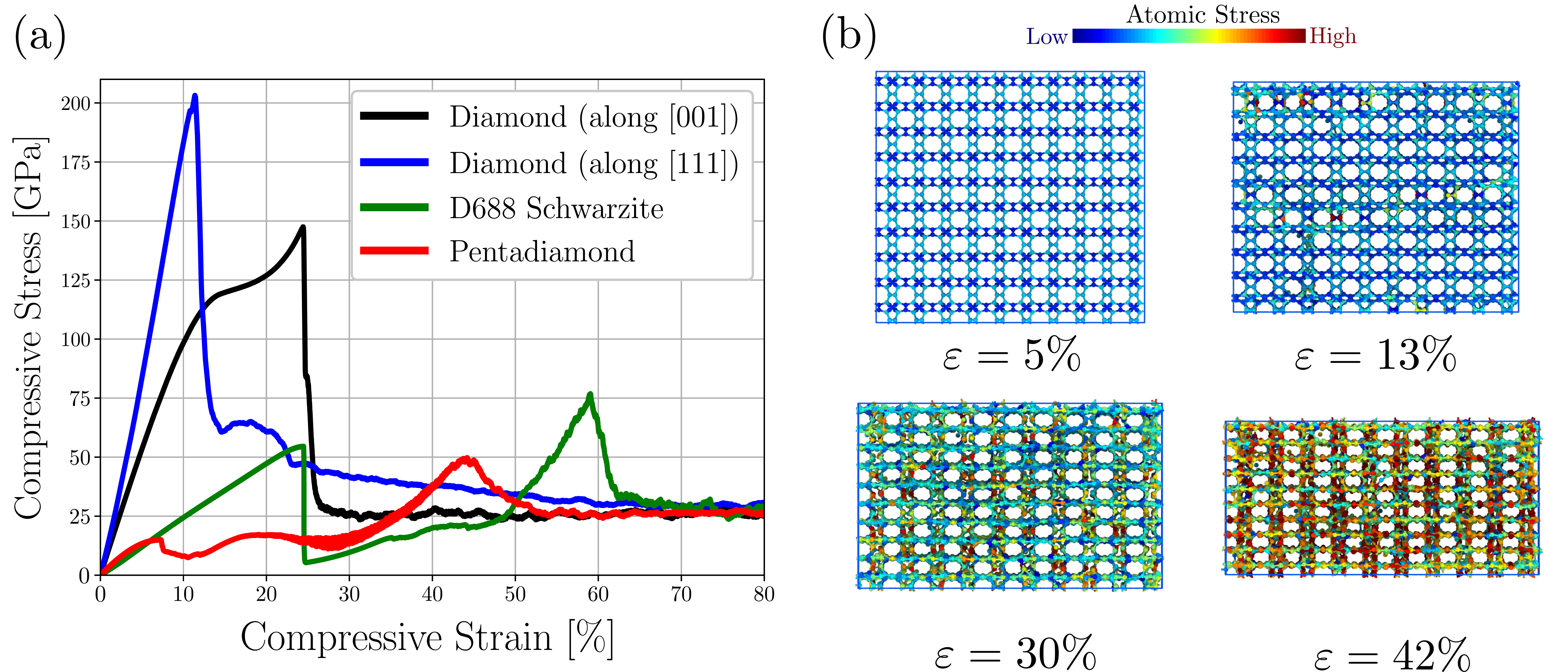}
    \caption{Compressive mechanical behavior from MD simulations. (a) Stress-strain curves of diamond along [001] (black), along [111] (blue), D688 schwarzite (green) and pentadiamond (red). (b) Representative MD snapshots at different strain values of the compression of pentadiamond. The color of atoms represent the von Mises stress values per atom calculated from equation \ref{eq:atomvonmises}.}
    \label{fig:compressive}
\end{figure}

All results reported on Table \ref{tab:structures+dft} and Figure \ref{fig:elfmaps} elucidate the mechanical behavior in the elastic regime, i.e., low strain values. In Figure \ref{fig:compressive} we present the MD results for compression of pentadiamond beyond the elastic regime. We compared the SS curves of diamond along [001] and [111] directions, D688 schwarzite and pentadiamond in Figure \ref{fig:compressive}~(a). For diamond, we observe an initial linear behavior for low strain values which is followed by a decrease of the slope that characterizes a plastic flow and, then, a maximum peak corresponding to the compressive strength is reached. After that, an abrupt drop occurs because failure (fracture) takes place. As expected, diamond is more resistant under mechanical compression along [111] (blue curve), than along the [001] one (black curve) since the stress values along [111] are larger. The shape of SS curves of D688 schwarzite and pentadiamond is slightly different as two and three stress peaks, respectively, are observed instead of just one for diamond. For D688 schwarzite, the first peak is associated with a structural transition from cubic to triclinic crystal system (See Supplementary Video S1) and the second peak corresponds to fracture. The first two peaks in the SS curve of pentadiamond does not have any association with changes in cubic symmetry, as can be seen in the Supplementary Video S2, but with fluctuations of the atomic positions spread out along the whole structure. The last peak corresponds to fracture. To further elucidate the stress distribution in pentadiamond we showed four successive MD snapshots in Figure \ref{fig:compressive}~(b). A non-uniform stress distribution occurs in which strong fluctuations of the atomic positions generate the instabilities showed in the SS curve of pentadiamond. In other words, the atomic stress on pentadiamond does not accumulate in specific spots of the structure, as occurs in the schwarzite (see the supplementary videos S1 and S2).

\begin{figure}[ht]
    \centering
    \includegraphics[scale=0.35]{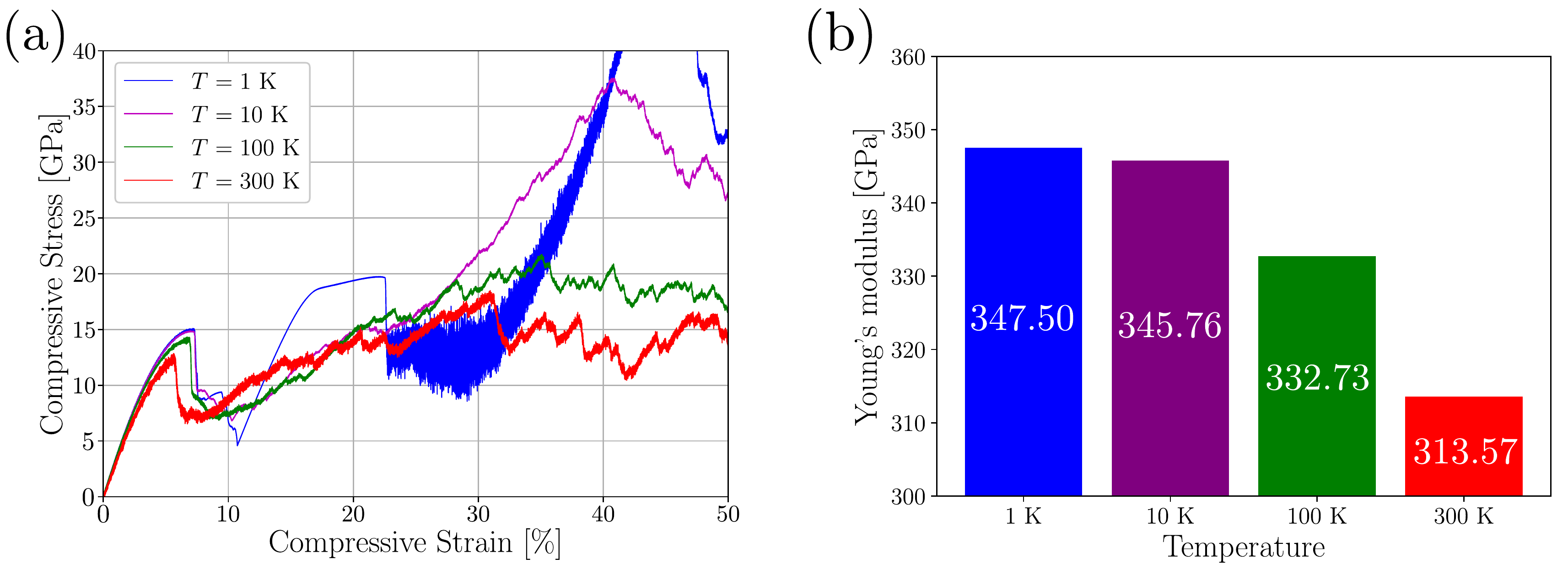}
    \caption{Temperature dependence on the elastic behavior of pentadiamond under compression. (a) Stress-strain curves corresponding to temperatures of 1 K (blue), 10 K (magenta), 100 K (green) and 300 K (red). (b) Graphic bars showing the Young's modulus values for different temperatures. }
    \label{fig:temperature}
\end{figure}

In order to verify the thermal stability of pentadiamond during compression we studied the temperature dependence on the SS curves and the Young's modulus, shown in Figure \ref{fig:temperature}~(a) and (b), respectively.
An interesting difference between the SS curves at 1 K and at higher temperatures, as shown in Figure \ref{fig:temperature}~(a), is the absence of a well defined curve (without fluctuations) between $\sim$11\% and $\sim$22\% strain shown in the blue curve. SS curves corresponding to the temperatures of 1 K and 10 K have three and two stress peaks, respectively, where the first peak indicates yielding and the last one appears when the structure breaks. For higher temperatures, such as 100 K and 300 K, we only observe a single stress peak that is followed by strong fluctuations that appear as the structures is being compressed. This happens because this single peak already characterizes failure and the material starts to break (See Supplementary Video S3). This indicates that the mechanical stability of pentadiamond under mechanical compression is strongly decreased at room temperature since the fracture strain goes from $\sim$44\% at 1 K (blue curve) to $\sim$5\% at 300 K (red curve). Although the failure point of pentadiamond is dramatically decreased, while the Young's modulus is only slightly decreased as the temperature increases, as can be seen from Figure \ref{fig:temperature}~(b), that shows a decrease of $\sim$9\% from its value at 1 K to room temperature. 

\begin{figure}[ht]
    \centering
    \includegraphics[scale=0.15]{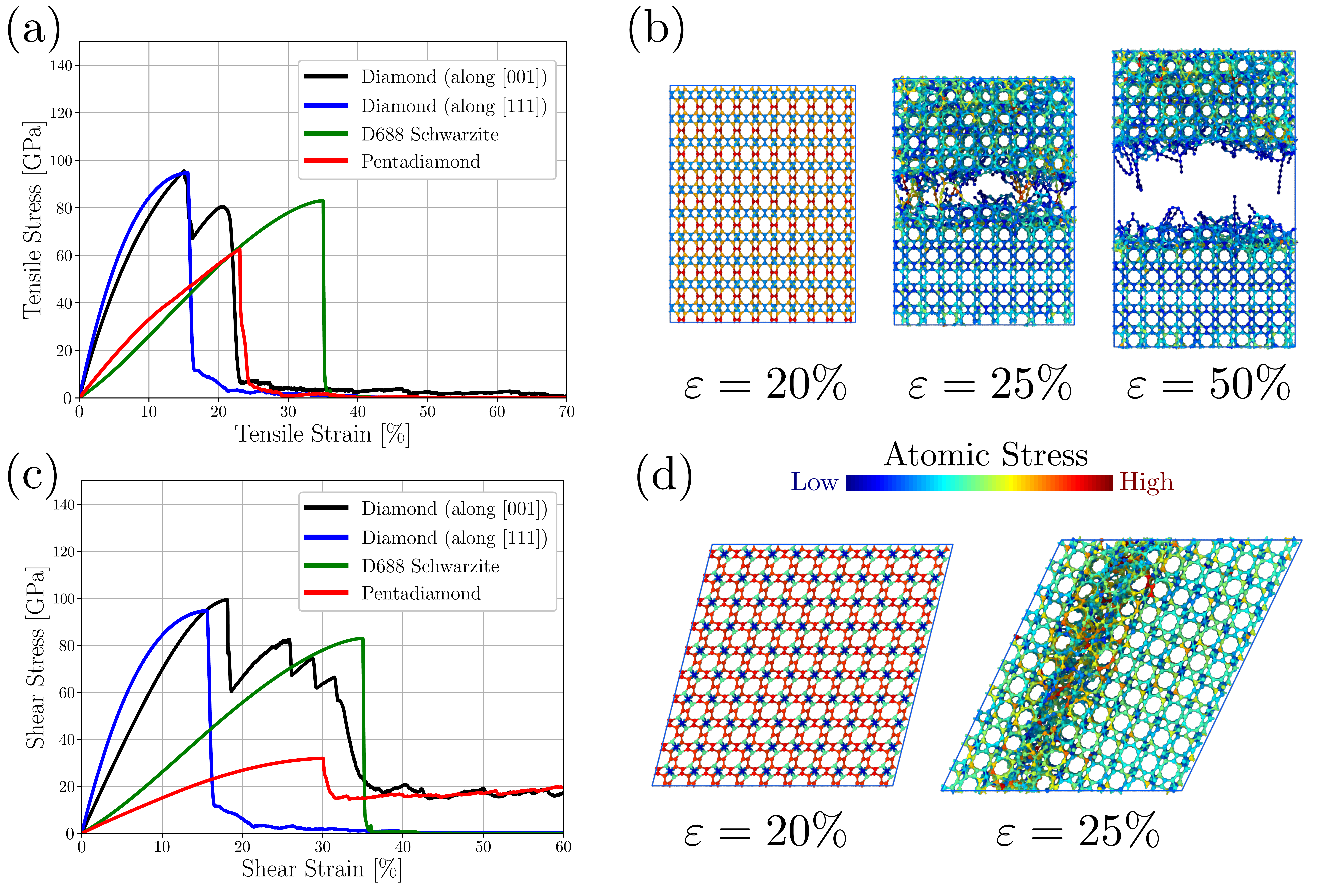}
    \caption{Tensile and shear deformation results from MD simulations. Tensile (a) SS curves and (b) two MD snapshots showing the von Mises stress profile. Shear (c) SS curves and (d) snapshots of the MD trajectory. For diamond, the black and blue curves correspond to the cases where the z axis corresponds to, respectively, [001] and [111]. The color of atoms are indicated on the color bar of the atomic stress.}
    \label{fig:tensile-shear}
\end{figure}

Tensile and shear deformation modes were also investigated in this work. In Figure \ref{fig:tensile-shear}~(a) we show the SS curves of diamond, schwarzite and pentadiamond. All curves begin with an approximately linear stress-strain dependence that is followed by plastic flow until a maximum peak is reach and, then, they abruptly drop to zero stress which indicates rupture (fracture). The only exception is the SS curve of diamond stretched along [001], where a second peak is observed due to internal shear fractures before it ultimately breaks (for more details see the Supplementary Video S4). MD snapshots at different stages of tensile deformation of pentadiamond are shown in Figure \ref{fig:tensile-shear}~(b) along with the atomic von Mises stress. At $\varepsilon=20\%$, right before it breaks, the structure possess a well layered stress distribution which indicates brittle fracture in a plane perpendicular to the direction of loading, which is subsequently confirmed at higher strains. For more details of tensile deformation see the Supplementary Video S5. For shear deformations, the SS curves (Figure \ref{fig:tensile-shear}~(c)) have similar shape as the tensile ones. Two of these curves, however, do not go to zero because of limitations on how large the tilt factor of the simulations box can be on LAMMPS, even by turning off the \texttt{flip} keyword on the \texttt{fix deform} command. This means that, at the maximum shear strain reach in the simulation, complete rupture was not reached yet. If we could shear the simulation box further to reach higher shear strains those SS curves would eventually drop to zero. The SS curve for diamond with [001] direction along the z axis (black curve in Figure \ref{fig:tensile-shear}~(c)) possess a saw-tooth shape that is characteristic in materials where plastic flow precedes rupture. In the Supplementary Video S6 we can observe failure along the slip planes of diamond before the abrupt drop to a constant value. The same layered stress distribution is observed for pentadiamond under shear deformation, shown in Figure \ref{fig:tensile-shear}~(d), but now the layers are diagonally displayed. Although we observe failure at $\varepsilon=25\%$, complete rupture is not observed in the pentadiamond structure (see Supplementary Video S7).

The elastic moduli, such as Young's modulus, stress and strain values at fracture are reported in Table \ref{tab:md-results} for the three deformation modes (compressive, tensile and shear) studied in this work. As expected, the Young's modulus ($Y$) values of diamond when compressed (or stretched) along the [111] direction are considerably higher than the corresponding values when the load is along the [001] direction. From our considerations about the ELF of diamond, schwarzite and pentadiamond we concluded that the compressibility of pentadiamond should be intermediate between those of diamond and schwarzite as the electrons in sp$^3$ bonds are more spatially localized than in sp$^2$ ones. These conclusions are further validated by our MD simulations for both compressive and tensile deformations. This is also true for tensile and shear deformations, showing that the equilibrium electron distribution can provide qualitative predictions on the mechanical properties of crystals~\cite{brazhkin_2019}. Although the stiffness of pentadiamond in the elastic regime is considerably higher than that of schwarzite, the latter possesses a higher strength for all three deformation modes, which could be attributed to the presence of the strong fluctuations of atomic positions in pentadiamond, as shown in Figure \ref{fig:compressive}~(b). Those fluctuations are responsible for lowering the stress as the strain increases beyond the elastic regime. It is important to emphasize that this decrease in stress results from a different mechanism than that for schwarzites, where we observed a plateau in SS curve due to the collapse of its porosity~\cite{woellner_2018,felix_2019,gong_2020}. D688 schwarzite has the highest fracture strain values for the three deformation modes studied in this work. Among all crystalline carbon materials, schwarzites remain the ones that can be compressed at higher strains before breaking~\cite{woellner_2018,felix_2019}.

\begin{table}[ht]
\caption{Mechanical properties of diamond, D688 schwarzite and pentadiamond from MD simulations. The cubic supercell size used for each structure and the corresponding number of atoms ($N$) are indicated. The Young's modulus for the compressive ($Y$(C)) and tensile ($Y$(T)) deformations are listed, as well as the shear modulus ($G$), alongside with their corresponding compressive ($CS$), tensile ($TS$) and shear ($SS$) strengths. Finally, the fracture strain of compressive ($\varepsilon_F$(C)), tensile ($\varepsilon_F$(T)) and shear ($\varepsilon_F$(S)) loads are listed.}
\label{tab:md-results}
\begin{center}
\resizebox{\columnwidth}{!}{%
\begin{tabular}{c|c|c|c|c|c|c|c|c|c|c|c}
\textbf{Structure} & Supercell size & $\boldsymbol{N}$ & $\boldsymbol{Y}$\textbf{(C) [GPa]} & $\boldsymbol{CS}$ \textbf{[GPa]} & $\boldsymbol{\varepsilon_F}$\textbf{(C) [\%]} & $\boldsymbol{Y}$\textbf{(T) [GPa]} & $\boldsymbol{TS}$ \textbf{[GPa]} & $\boldsymbol{\varepsilon_F}$\textbf{(T) [\%]} & $\boldsymbol{G}$\textbf{ [GPa]} & $\boldsymbol{SS}$ \textbf{[GPa]} & $\boldsymbol{\varepsilon_F}$\textbf{(S) [\%]} \\ \hline
Diamond (along [001]) & $8\times 8\times 8$ & 4096 & 1092.39 & 147.55 & 24.42 & 1055.32 & 95.51 & 15.04 & 751.79 & 99.49 & 18.11 \\ \hline
Diamond (along [111]) & $8\times 8\times 8$ & 4096 & 1441.40 & 203.20 & 11.38 & 1319.90 & 94.75 & 15.60 & 596.49 & 94.76 & 15.59 \\ \hline
D688 Schwarzite & $6\times 6\times 6$ & 5184 & 211.59 & 76.83 & 59.03 & 203.90 & 82.95 & 35.01 & 145.83 & 82.95 & 35.00 \\ \hline
Pentadiamond & $5\times 5\times 5$  & 11000 & 347.50 & 49.77 & 44.19 & 362.98 & 62.67 & 23.05 & 163.31 & 31.92 & 30.00 \\ \hline
\end{tabular}
}
\end{center}
\end{table}

\section{Conclusions}

The mechanical and electronic properites, as well as the thermal stability of pentadiamond were investigated using DFT and Molecular Dynamics (MD) simulations. Since pentadiamond is a material where both sp$^2$ and sp$^3$ carbon hybridizations, for comparison purposes we also investigated the behavior of diamond and D688 schwarzite that are fully sp$^3$ and sp$^2$ hybridized materials, respectively. 

The electron distribution within the covalent bonds in pentadiamond shows a mixing of sp$^2$ and sp$^3$, as expected, and corroborates the fact that its mechanical strength should be intermediate between that of schwarzite and diamond. This is important to confirm the relation between electron localization and stiffness of crystalline materials~\cite{brazhkin_2019}. 

We also investigated the mechanical behavior of pentadiamond beyond the elastic regime by fully atomistic reactive MD simulations. The compressive behavior of pentadiamond is distinct from diamond and schwarzite because of the strong fluctuations of the atomic positions after yield strain, which characterizes a plastic flow. In contrast to D688 schwarzite, pentadiamond does not exhibit a cubic to triclinic transition. 

As we increase temperature, as expected, Young's modulus values decrease, but this variation (up to 300 K) is smaller than 10\% (from 347.5 to 313.6 GPa), which is approximately one third of the corresponding diamond value.but this variation (up to 300 K) is smaller than 10\% (from 347.5 to 313.6 GPa). However, the fracture strain is very sensitive, varying from $\sim$44\% at 1K to $\sim$5\% at 300K.

In summary, this new carbon allotrope (pentadiamond) exhibits some interesting mechanical and electronic properties. Some of its behavior are intermediate between diamond (pure sp$^3$) and schwarzite (pure sp$^2$) and it a direct consequence of having mixed sp$^2$ and sp$^3$ carbon bonds.

\section*{Acknowledgements}
This work was financed in part by the Coordenacão de Aperfeiçoamento de Pessoal de Nível Superior - Brasil (CAPES) - Finance Code 001, CNPq, and FAPESP. We thank the Center for Computing in Engineering and Sciences at Unicamp for financial support through the FAPESP/CEPID Grants \#2013/08293-7 and \#2018/11352-7. We also thank Conselho Nacional de Desenvolvimento Cientifico e Tecnológico (CNPq) for their financial support.

\bibliography{mybibfile}

\end{document}